\documentclass[useAMS,usenatbib]{mn2e}
\usepackage{graphicx,amssym}
\def\<{\langle}
\def\>{\rangle}
\def\lsim{\mathrel{\hbox{\rlap{\hbox{\lower4pt\hbox{$\sim$}}}\hbox{$<$}}}}
\def\gsim{\mathrel{\hbox{\rlap{\hbox{\lower4pt\hbox{$\sim$}}}\hbox{$>$}}}}

\newcommand{\bc}{\begin{center}}
\newcommand{\ec}{\end{center}}

\title[Discreteness effects in HDM/WDM simulations]
      {Discreteness Effects in Simulations of Hot/Warm Dark Matter}

\author[J. ~Wang \& S. D. M. ~White ]
       {Jie Wang$$\thanks{Email: wangjie@mpa-garching.mpg.de},
        Simon D.~M.~White$$ \\
        $^1$Max--Planck--Institut f\"ur Astrophysik,
        Karl--Schwarzschild--Str. 1, D-85748 Garching, Germany\\
        }
\begin{document}

\date{Accepted 2007 May 25.
      Received 2006 May 25;
      in original form 2007 February 21}

\pagerange{\pageref{firstpage}--\pageref{lastpage}}
\pubyear{2007}

\maketitle

\label{firstpage}

\begin{abstract}
  In Hot or Warm Dark Matter universes the density fluctuations at early times
  contain very little power below a characteristic wavelength related
  inversely to the particle mass. We study how discreteness noise influences
  the growth of nonlinear structures smaller than this coherence scale in
  $N$-body simulations of cosmic structure formation. It has been known for 20
  years that HDM simulations in which the initial uniform particle load is a
  cubic lattice exhibit artifacts related to this lattice.  In particular, the
  filaments which form in such simulations break up into regularly spaced
  clumps which reflect the initial grid pattern. We demonstrate that a similar
  artifact is present even when the initial uniform particle load is not a
  lattice, but rather a glass with no preferred directions and no long-range
  coherence. Such regular fragmentation also occurs in simulations of the
  collapse of idealised, uniform filaments, although not in simulations of the
  collapse of infinite uniform sheets. In HDM or WDM simulations all
  self-bound nonlinear structures with masses much smaller than the free 
  streaming mass appear to originate through spurious fragmentation of
  filaments. These artificial fragments form below a characteristic mass which
  scales as $m_p^{1/3}k_{peak}^{-2}$, where $m_p$ is the $N$-body particle
  mass and $k_{peak}$ is the wavenumber at the maximum of $k^3 P(k)$
  ($P(k)$ is the power spectrum). This has the unfortunate
  consequence that the effective mass resolution of such simulations 
  improves only as the cube root of the number of particles employed.
\end{abstract}

\begin{keywords}
 methods: N-body simulations --methods: numerical -- dark matter:
 massive neutrinos
\end{keywords}

%%%%%%%%%%%%%%%%%%%%%%%%%%%%%%%%%%%%%%%%%%%%%%%%%%%%%%%%%%%%%%%
\section{Introduction}
\label{sec:intro}
In the absence of a full analytic understanding of nonlinear structure growth,
numerical simulations provide a critical link between the weak density
fluctuations measured in the cosmic microwave background and the strong
inhomogeneities observed on all but the very largest scales in the present
Universe. Indeed, numerical simulations played a decisive role in excluding
massive neutrinos as a dark matter candidate \citep{white83} and
in establishing the $\Lambda$CDM model as the leading and now standard
paradigm for the formation of all structure \citep{davis85,white87,cen94,navarro96}. 
With the development of
more powerful computer hardware, of more accurate numerical algorithms, and of
methods to follow additional physical processes, the importance of simulations
as a tool to interpret observations of observed structure continues to
increase dramatically.  In this paper we are concerned with one aspect of the
simplest kind of cosmological structure formation simulation, namely how
discreteness effects can drive the growth of spurious small-scale structure in
N-body simulations of evolution from initial conditions containing no such
structure.
% Grid & Glass

To create initial conditions for a cosmological simulation, a uniform particle
distribution is needed. This can be perturbed by a random realisation of the
linear fluctuation field associated with the specific structure formation
model to be simulated (e.g.  $\Lambda$CDM). A uniform Poisson distribution of
particle positions is not suitable for this purpose, because stochastic
``root-$N$'' fluctuations can exceed the density fluctuations predicted by the
desired model over a wide range of scales. To avoid this problem, most early
simulations chose a regular cubic lattice as the initial uniform load.
Symmetry then assures that there can be no growth of structure in the absence
of imposed perturbations \citep{efstathiou85}. The preferred
directions and the large-scale coherence of the lattice may, however, be a
disadvantage, since they can give rise to numerical artifacts. As an
alternative, \citet{white96} suggested using a glass-like initial particle load
created by carrying out a cosmological simulation from Poisson initial
conditions but with the sign of the peculiar gravitational accelerations
reversed, so that each particle is repelled by all the others.  When such a
system reaches quasi-equilibrium, the total force on each particle vanishes,
as for a grid, but there are no preferred directions and no long-range
order. The power spectrum on scales much larger than the mean interparticle
spacing approaches a power-law $P(k)\sim k^{n}$ with $n=4$ (Baugh et
al. 1995), where $n=4$ is the minimal large-scale power expected for a
discrete stochastic system \citep[][section 28]{peebles80}.

Normally, grid and glass initial loads are considered equivalent.
Nevertheless, artifacts due the initial lattice are obvious in early images of
the sheets, filaments and ``voids'' formed in Hot Dark Matter (HDM)
simulations \citep[e.g.][]{Centrella83,frenk84,efstathiou85,
centrella88}. \citet{Baugh95} and \citet{white96}
showed that low density regions appear very different in simulations
with a grid initial load than in simulations started from a glass, although
Baugh et al. found that this does not show up as a difference in their power
spectra. Nevertheless, the regularly spaced clumps seen along filaments in HDM
simulations are clearly related to the initial particle grid, and so seem
unlikely to reflect a true physical instability.  Despite this, \citet{bode01}
and \citet{knebe03} interpreted analogous structures in their Warm Dark Matter
(WDM) simulations (which were set up using a grid initial load) as the result
of the physical fragmentation of filaments. The nonlinear formation
of such small-scale structure could have important consequences in models like
HDM or WDM where power on small scales is strongly suppressed in the linear
initial conditions.  It is thus important to establish which simulated
structures are real and which are artifacts, as well as to understand whether
the simulations can be improved by, for example, choosing a glass initial load
in place of a grid.

% filament fragmentation in \lambdaWDM
\citet{gotz02,gotz03} carried out WDM simulations using both
grid and glass initial loads and reported significant differences. With a grid
they found spurious low-mass halos evenly spaced along filaments, exactly as
in earlier HDM experiments.  The spacing is simply that of the initial grid,
stretched or compressed by the large-scale distortion field.  They emphasised,
however, that such unphysical halos were less evident in their simulations
starting from a glass. This conclusion disagrees with our own work below,
where we find spurious halos also in simulations from glass initial conditions
and with a frequency very similar to that found in the grid case. Curiously,
in the glass case also we find the spurious halos to be regularly spaced along
filaments even though the initial condition is \emph{not} regular over the
relevant scales.

In this paper, we wish clarify this issue by isolating the numerical
artifact, by exhibiting it in idealised filament formation
simulations, by exploring its dependence on the nature of the uniform
particle load, and by establishing the dependence of its
characteristic scale on the discreteness scale of the simulation and
the coherence scale of the WDM/HDM initial conditions. We carry out
cosmological simulations of an HDM universe at a wide range of
resolutions and with both grid and glass initial loads. In addition,
we simulate the collapse of an infinite straight uniform density
filament from glass initial conditions, showing that it fragments into
regularly spaced clumps. To gain additional insight, we also consider
the collapse of a glass to a uniform sheet, and the growth of
structure in a uniform, space-filling, but anisotropically compressed
glass. Rapid fragmentation on small scales occurs only in the filament
case. Our tests also demonstrate that considerable care is needed to
produce an initial glass load for which the growth of small-scale
structure in filaments is optimally suppressed. We propose a
randomisation technique which successfully washes out most
code-dependent periodic signals in the initial load.

The remainder of our paper is organised as follows. In $\S2$ we first discuss
the aspects of our simulation code which are relevant to the problem at hand,
in particular, how it estimates gravitational accelerations and how it is
modified in order to create a uniform glass distribution.  We then describe
the way in which initial conditions are created for the simulations presented
in the rest of the paper.  $\S3$ presents results from our Hot Dark Matter
simulations, showing that all small-scale collapsed structures appear to form
initially as regularly spaced clumps along filaments, and that these are
similar for grid and for glass initial loads. Results for our studies of
idealised structure formation from anisotropically compressed glasses are
presented in $\S4$. Rapid fragmentation on small scales occurs only in the
filament case. $\S5$ examines this filament fragmentation in more detail,
showing that its characteristic scale is related to the interparticle
separation for a well-constructed glass, but that scales related to the
Poisson-solver of the glass-construction code can play an important role if
their influence is not carefully controlled. Finally, $\S6$ summarises the
implications of our results for simulations of structure formation. In
particular, we show that for nonlinear structures the effective mass
resolution of simulations of HDM or WDM universes improves only as the cube
root of the number of simulation particles employed. This is much more
pessimistic than the direct proportionality to $N$ which might naively have
been expected.
%%%%%%%%%%%%%%%%%%%%%%%%%%%%%%%%%%%%%%%%%%%%%%%%%%%%%%%%%%%%%%%%%%%%%%%%%%%%%

\section {Simulation methods and initial conditions}
\label{sec:simulation}

All simulations in this paper were performed using the massively parallel
N-body code L-Gadget2. This is a lean version of Gadget2 \citep{springel05a}
with
the SPH part excluded and with the memory requirements minimised. It was
originally written in order to carry out the \emph{Millennium Simulation}
\citep{springel05b}.

The computation of gravitational forces is the most critical and
time-consuming element of any cosmological $N$-body code. Gadget2 uses a
hybrid tree-PM method where the long-range force is calculated at low
resolution using a particle-mesh scheme, and is supplemented by a
high-resolution but short-range correction calculated using a tree algorithm.
The short-range correction is assembled in real space by collecting
contributions from all neighbouring particles. The long-range force is
calculated by assigning the particles to a regular cubic mesh, by using
Fourier methods to obtain the corresponding potential, and by numerically
differencing the result.  For a single particle this scheme introduces a
maximum force error of 1-2 percent near the split scale. Choosing a suitable
split scale (typically several times the mean interparticle separation)
results in force errors for smooth distributions of particles which are almost
everywhere far below 1 percent. Gadget2 uses a space-filling fractal, the
Peano-Hilbert curve, to control the domain decomposition associated with
parallelisation. Because there is a good correspondence between the spatial
decomposition obtained from this self-similar curve and the hierarchical tree
used to compute forces, it is possible to ensure that the tree decomposition
used by the code is independent of the platform, in particular of the number
of processors on which it is run.  In addition, the ``round-off'' errors in
the forces induced when summing contributions from all processors are
explicitly considered. As a result, the forces are independent of
the number of processors and the domain cuts that are made. We believe that
all code-related numerical effects relating to the calculation of
gravitational accelerations are well under control in Gadget2.

Glass construction is embedded in Gadget2 by using some compile options. A
preset number of particles is initially distributed at random within the
cubic computational volume and the standard scheme is used to obtain the
gravitational acceleration on every particle. After reversing the signs of the
accelerations, all particles are advanced for a suitably chosen timestep. The
velocities are then reset to zero and the whole procedure is repeated.  After
about a hundred steps the acceleration of each particle approaches zero.
Notice that this is the acceleration as obtained by the code, including the
effects of force anisotropy, domain decomposition, etc.  Because the glass is
made in a periodic cube, we can get a large glass file cheaply by tiling a big
box with many replications of the original glass.  However, the accelerations
calculated by the code may no longer vanish exactly for this larger glass
because the force anisotropies and inaccuracies now occur on a different scale
than when the glass was created.  

The FFT calculation and the Barnes \& Hut tree used in Gadget2 are
both based on static grids.  This \emph{spatially fixed} decomposition
introduces weak periodic signals in the force calculation, and these
are reflected in the particle distribution at the end of the
glass-making procedure. We will see below that that this can introduce
measurable spikes in the 1-D power spectrum of the final glass.  To
reduce such effects we randomly offset the particle distribution in
all three coordinates with respect to the computational box before
carrying out each force computation during glass-making. This
suppresses the induced signals quite effectively but does not fully
eliminate them.  A glass constructed in this fashion is referred to as
a ``good'' glass in the following, while a glass constructed without
the random offset technique and showing significant high spikes is
referred to as a ``poor'' glass. In the rest of this paper, we use
``good'' glasses for our initial conditions except where explicitly
noted. These issues are further discussed in section 5.

Below we consider grid initial loads in addition to glasses in order to
compare their performance and to check the results of previous work
\citep{bode01,gotz02,gotz03}.  A third quasi-uniform particle
distribution, the Quaquaversal distribution, has recently been suggested by
\citet{hansen07} as a non-periodic uniform initial load, a possible
alternative to a glass.  We have created such a Quaquaversal load with
$2\times8^7$ particles using the code provided by Hansen et al at their web
site, and we compare its performance to our grid and glass initial loads in
an Appendix.  It produces significantly worse discreteness artifacts than
either grids or glasses.

Two types of simulation are considered below. The first is a series of
cosmological simulations of evolution from Hot Dark Matter (HDM)
initial conditions. Most of these are for a single realisation of the
HDM density field within a 100$h^{-1}$Mpc cube, but with different
kinds of initial load and with varying mass resolution. One considers
a 200$h^{-1}$Mpc cube in order to better constrain the abundance of
large objects. For these simulations, we choose an Einstein-de Sitter
universe dominated by a single massive neutrino. We take $H_0=76.5
{\rm km s^{-1}Mpc^{-1}}$ which implies a neutrino mass of $\sim 55
{\rm eV}$ and a corresponding free-streaming scale $\lambda= 22.2~{\rm
Mpc}$ \citep{bond83} below which initial fluctuations are
exponentially suppressed relative to an assumed $P(k)\propto k$
primordial power spectrum. The power spectrum we actually use to
impose fluctuations on our initial loads is based on the theoretical
predictions of \citet{bardeen86} and agrees with numerical estimates
from the Boltzmann solver CMBFAST \citep{Seljak96}.  Since the same
realisation is used for all our 100$h^{-1}$Mpc simulations, they
should all produce identical structures. We start integrating at
redshift $z=20$ and evolve structure to a formal present-day amplitude
of $\sigma_8=2$. This corresponds to the collapse of the first
nonlinear structure in the simulation at $z \sim 6$. As a check of our
starting redshift we reran one simulation glass128 starting from
$z=100$. At $z=15$ the power spectrum of this simulation differed from
that of the original run by few percent or less on all scales.

Our simulations are listed with their parameters in Table 1: pre-IC, L, $m_p$,
$\epsilon$,and $n_p$ denote the initial load, the box size, the particle mass,
the softening length, and the particle number respectively.
\begin{table*}
\caption{HDM simulations}
\begin{center}
\begin{tabular}{cccccccccc} \hline
name&$pre-IC$ &$L [h^{-1}{\rm Mpc}]$ & $m_p[h^{-1}M_\odot]$ &
   $\epsilon (h^{-1}{\rm Mpc})$ & $n_p$ & \\ \hline
glass64 & glass &100 & $1.06\times10^{12}$ & 0.08 & $64^3$   \\
glass128& glass &100 & $1.32\times10^{11}$ & 0.04 & $128^3$   \\
glass256 & glass & 100 & $1.65\times10^{10}$ & 0.02 & $256^3$  \\
glass256-2 & glass & 200 & $1.32\times10^{11}$ & 0.04 & $256^3$  \\
glass512 & glass &100 & $2.07\times10^9$ & 0.01 & $512^3$ & \\
grid128 & grid  &100 & $1.32\times10^{11}$ & 0.04 & $128^3$ \\
qset134 & Q-set &100 & $1.32\times10^{11}$ & 0.04 & $\sim134^3$ \\
\hline
 \end{tabular}
\end{center} \end{table*}

%%%%%%%%%%%%%%%%%%%%%%%%%%%%%%%%%%%%%%%%%%%%%%%%%%%%%%%%%%%%%%%%%%%%%%%%%
%%%%%%%%%%%%%%%%%%%%%%%%%%%%%%%%%%%%%%%%%%%%%%%%%%%%%%%%%%%%%%%%%%%%%%%%%
Our second type of simulation is designed specifically to study the
discreteness artifacts which show up in the filamentary structures 
within our HDM simulations. These simulations follow evolution from a variety
of highly idealised initial conditions, all based on uniformly but
anisotropically compressed glasses. We consider three different cases:

\emph{Anisotropic glass}: The glass initial load is compressed by a factor of
2 along one axis, by a factor of 3 along a second axis, and is unaltered along
the third axis. Six replications of this configuration are then used to tile
the computational cube to produce an initial condition which is uniform and
glass-like on large scales but where the forces are no longer balanced on the
scale of the interparticle separation.

\emph{Sheet}: The glass initial load is compressed along one dimension by a
factor of 2, so that it fills half of the computational volume. The other half
remains empty.  This configuration collapses to form an infinite uniform
sheet.

\emph{Filament}: Our glass initial load is compressed by a factor of 2 along
two of its periodic directions while the third remains unchanged.  The
particles then fill a quarter of the computational volume, the rest remaining
empty. This configuration collapses to form a uniform straight filament.

All the simulations carried out from these initial conditions assume an
Einstein-de Sitter background universe.  We define the expansion factor $a$ to
be unity at the initial time.

We identify collapsed ``halos'' in our HDM simulations using a
Friends-of-Friends (FOF) algorithm with linking length 0.2 times the mean
interparticle spacing (Davis et al 1985). In the following we will only
consider FOF halos with 32 or more particles. Subhalos within these halos were
identified using the SUBFIND algorithm \citep{springel01} with parameters
set to retain all overdense self-bound regions with at least 20 particles.
Based on these subhalo catalogues we use the techniques of \citet{springel05b} 
to construct merging trees which allow us to follow the formation
and evolution of all halos and subhalos.
%%%%%%%%%%%%%%%%%%%%%%%%%%%%%%%%%%%%%%%%%%%%%%%%%%%%%%%%%%%%%%%%%%%%%%%%%%%%%%
%%%%%%%%%figure 1%%%%%%%%
\begin{figure}
\bc
\hspace{-1.4cm}
\resizebox{8cm}{!}{\includegraphics{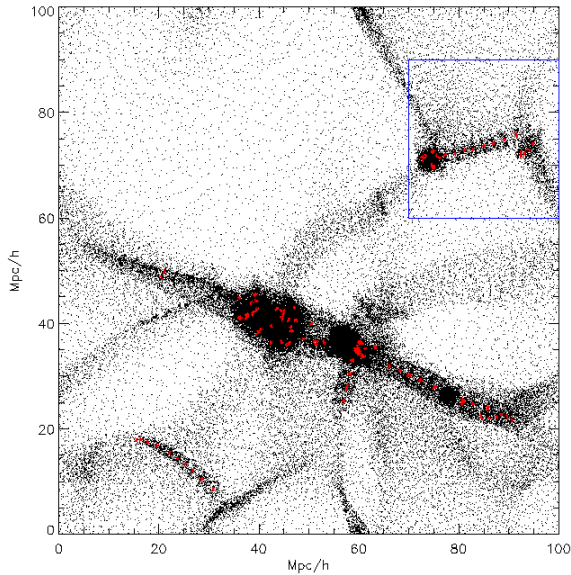}}\\%
\caption{ A slice of thickness $10h^{-1}{\rm Mpc}$ through the HDM simulation
glass128 at $z=0$. Small black points indicate individual simulation
particles, while the larger red points show FOF halos with $32\leq N_{fof} <
300$.
\label{fig1}}
\ec
\end{figure}

%%%%%%%%%figure 2%%%%%%%%%%%%%
\begin{figure}
\bc
\hspace{-1.4cm}
\resizebox{8cm}{!}{\includegraphics{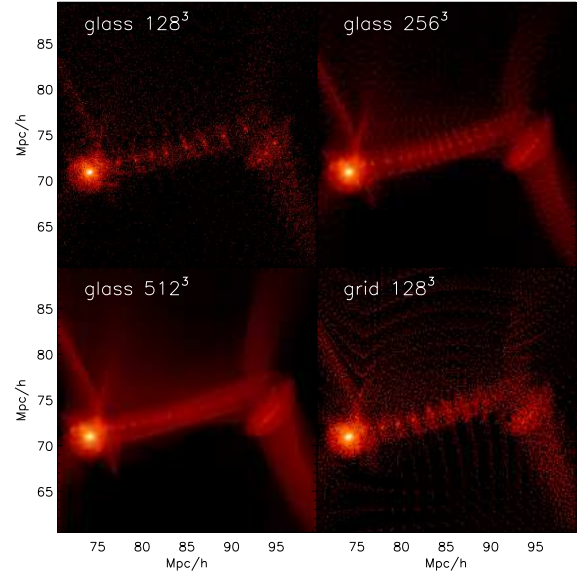}}\\%
\caption{ A subregion containing a filament at $z=0$ in four of our $100
h^{-1}$Mpc HDM simulations based on differing initial loads and differing mass
resolution.  The first three panels are taken from glass$128$, glass$256$ and
glass$512$ respectively, while the last one ( bottom right) is taken from
grid$128$.  The region shown here corresponds to that surrounded by a blue
square in Fig 1.
\label{fig2}}
\ec
\end{figure}

%%%%%%%%%figure 3%%%%%%%%%%%%%
\begin{figure*}
\bc
\hspace{-1.6cm}
\resizebox{17cm}{!}{\includegraphics{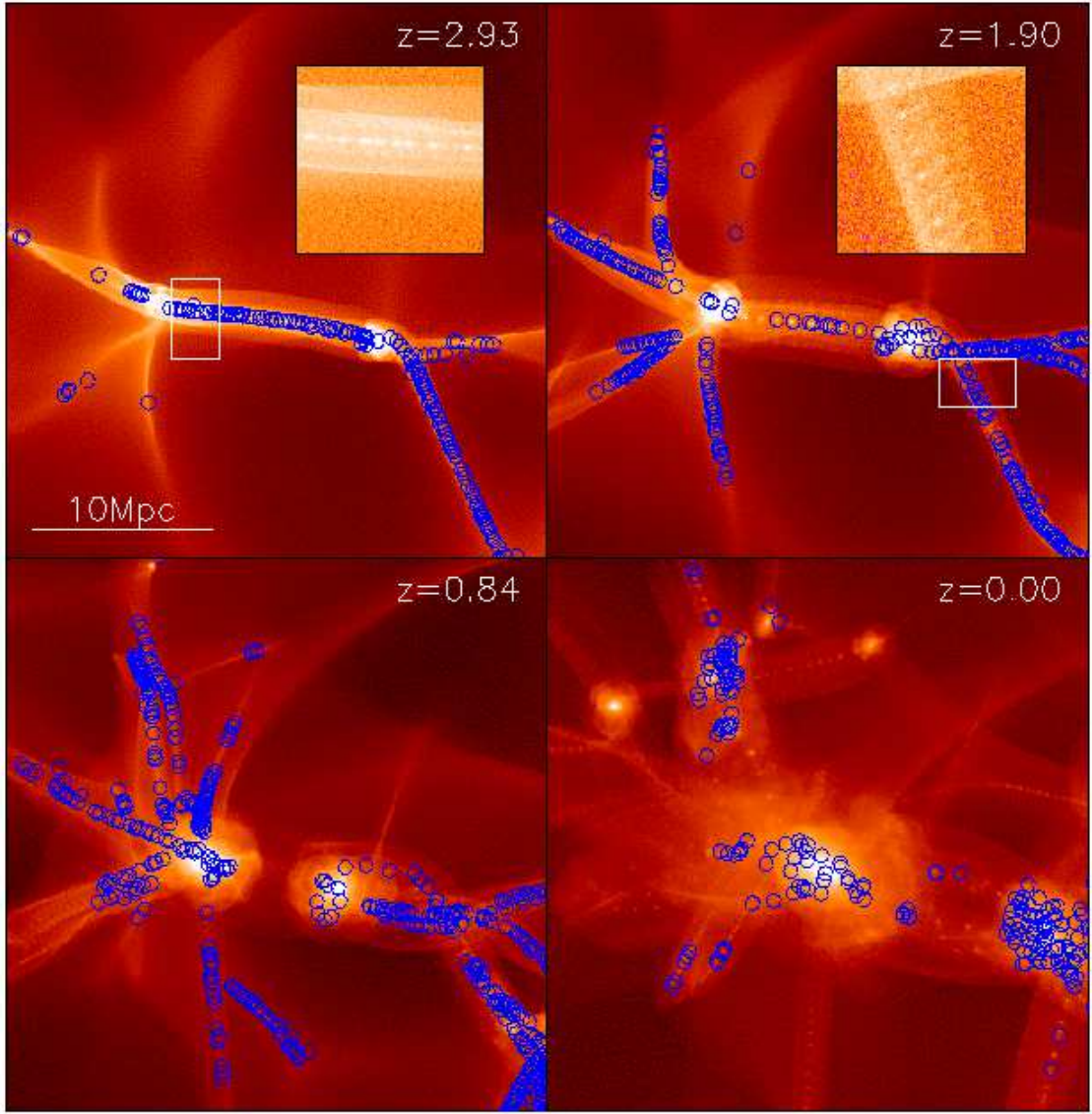}}\\%
\caption{The bottom right panel shows a subregion of glass512 at $z=0$
containing the largest FOF halo. The blue circles indicate all subhaloes with
more than 20 particles within this object. The other three panels show the
same subregion at earlier times ($z=0.84; 1.90; 2.93$) with the blue circles
now indicating the progenitors of all subhaloes identified at $z=0$.
It is striking that almost all these progenitors lie in 1-D
structures --- filaments --- at the earlier times. The two smaller plots in the
upper panels are stretched and magnified images of the filaments indicated by
white rectangles in the main panels.
\label{fig3}}
\ec
\end{figure*}

%%%%%%%%%%%%%%%%%%%%%%%%%%%%%%%%%%%%%%%%%%%%%%%%%%%%%%%%%%%%%%%%%%%%%%%%%%
\section {Filament Fragmentation in HDM Simulations}
\label{sec:filament_hdm}

The original motivation for this paper came from an unexpected phenomenon in
our HDM simulations. Even when we use a glass initial load, we find that the
filaments in these simulations break up into regularly spaced clumps, just as
in early simulations based on grid initial loads. We illustrate this in Fig.~1
which shows a slice through simulation glass128. All FOF haloes with $32\leq
N_{fof} < 300$ are indicated by red points.  It is obvious that many of these
low-mass haloes lie in the filaments, and that they are surprisingly regularly
spaced along them. 

We find similar behaviour in all our HDM simulations, independent of the
initial load and the mass resolution. In Fig.~2 we focus on a small cubic
subregion containing a filament. (This region is indicated by a blue square in
Fig.~1). In all four simulations, small clumps are visible at regularly spaced
positions along the filament. Their spacing is very similar in the two $128^3$
simulations, even though one started from a glass and the other from a grid.
The spacing is reduced by about a factor of two in the $256^3$ simulation and
by about another factor of two in the $512^3$ simulation. The clumps are
difficult to see in this last case, but this is merely a consequence of the
resolution of the image (compare Fig.~3 below). There are strong indications
that these clumps are a numerical artifact, not least because in the grid128
case they line up with the distorted but still recognisable pattern of the
initial grid, just as in early HDM simulations (e.g. Centrella \& Melott 1983)
or in the WDM simulations of \citet{gotz03}. \citet{gotz03} reported 
that the effect is absent when starting from a glass, although 
their Fig. 1 appears to show it in the filament at the lower right 
corner of their image.  \citet{bode01} also noticed that many low-mass 
haloes formed in the filaments of their WDM simulation, but they 
interpreted this as a result of physical pancake fragmentation.  
The tight relation between clump scale and mass resolution
manifest in Fig.~2 makes it clear, however, that this is actually a reflection
of N-body discreteness effects. Its surprising aspect is that the regular clump
spacing persists in the glass case where the initial load has no large-scale
coherence. We consider this issue further below.

Another consequence of this artifact is illustrated in Fig.~3. It appears that
almost all the subhaloes within the massive objects present at $z=0$ actually
originated through spurious filament fragmentation.  Fig. 3 focusses on a
small subregion of the glass512 simulation which contains the largest FOF
halo, an object with several linked density centres at $z=0$. The lower right
panel shows the final mass distribution in this cubic region. All subhaloes
with SUBFIND particle count greater than 20 are indicated by blue circles. The
other three panels use our merging trees to trace all progenitors of
these subhaloes back to earlier times. All of them appear to form initially as
evenly spaced ``beads'' strung along filaments. They later fall into the large
halo where they are seen at $z=0$. The artificial regularity of their
formation is illustrated by the two zooms in the upper panels. We conclude
that the first generations of haloes in pure HDM or WDM universes should
contain no dark matter subhaloes of smaller mass scale.

%%%%%%%%%%%%%%%%%%%%%%%%%%%%%%%%%%%%%%%%%%%%%%%%%%%%%%%%%%%%%%%%%%%%
%%%%%%%%%%%%%%%%%%%%%%%%%%%%%%%%%%%%%%%%%%%%%%%%%%%%%%%%%%%%%%%%%%%%
\section {Structure Growth in Idealised Glass Collapses}
\label{sec:2d}

In our HDM simulations there is very little power in the imposed HDM power
spectrum below the free-streaming scale. The regular fragmentation of the
filaments is clearly related to the mesh for a grid initial load, but the
origin of the equally regular fragmentation in the glass case is less
obvious. In Fig.4 we analyse the structure of a ``good'' $160^3$ particle
glass to search for signs of unexpected periodic behaviour.  The irregular
blue line in this figure is the dimensionless 3-D power per unit $\ln k$,
$\Delta^2(k) = k^3 P_3(k)$ where $P_3(k)$ is the 3-D power spectrum. For
comparison, the straight blue line gives the expectation for a Poisson
distribution with the same number of particles ($P(k)=1/N$). On large scales
the power in the glass is far below this white noise level, with
$\Delta^2(k)\propto k^7$ rather than $\Delta^2(k)\propto k^3$. For a relatively
narrow frequency band near $k=160$, however, the power is noticeably above the
Poisson expectation. This corresponds approximately to the
separation of the clumps which form on the filaments so there may be some
connection to this artifact in our HDM simulations.

Fig.~4 also shows the power per unit $\ln k$ in 2-D and 1-D projections of
this same glass, $\Delta^2(k) = k^2 P_2(k)$ and $\Delta^2(k) = k P_1(k)$
respectively. Again the measured results are compared to the expectation for a
Poisson distribution with the same number of particles. Both cases show
features directly analogous to those seen in the 3-D power spectrum. On large
scales (small $k$) the power is strongly suppressed relative to the white
noise level, with a spectrum which is steeper than white noise by four powers
of $k$. Near $k=160$ there is a narrow range of wavenumbers where the power
rises significantly above the white noise level. Again this feature could be
related to the break up of sheets or filaments into clumps spaced regularly at
about the mean interparticle separation.

%%%%%%%%%figure 4%%%%%%%%%%%%%
\begin{figure}
\bc
\hspace{-1.4cm}
\resizebox{8cm}{!}{\includegraphics{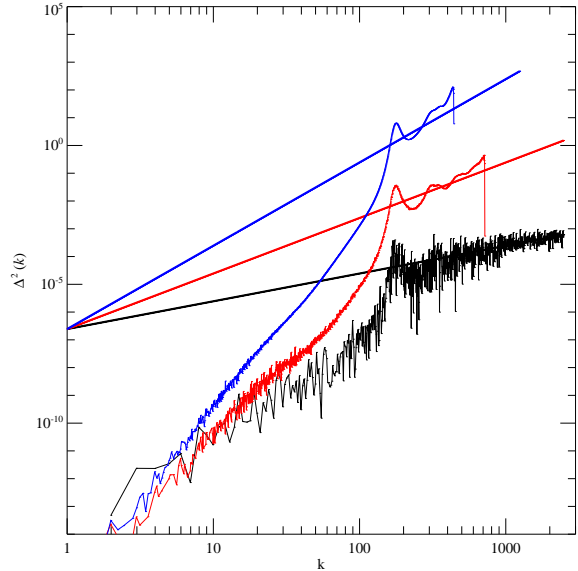}}\\%
\caption{ The 1-D (black), 2-D (red) and 3-D (blue) dimensionless power
per unit $\ln k$ for a glass with $160^3$ particles.  The straight lines are
the corresponding expectations for a Poisson distribution of $160^3$
particles. The wavenumber $k$ is given in natural units so that k=1 corresponds
to the fundamental mode of the computational cube. The glass used here is a
``good'' glass, where random position offsets at each timestep were used to
suppress artifacts due to anisotropies in the Gadget2 Poisson solver (see
section 5).
\label{fig4}}
\ec
\end{figure}

%%%%%%%%%figure 5%%%%%%%%%%%%%
\begin{figure}
\bc
\hspace{-1.4cm}
\resizebox{8cm}{!}{\includegraphics{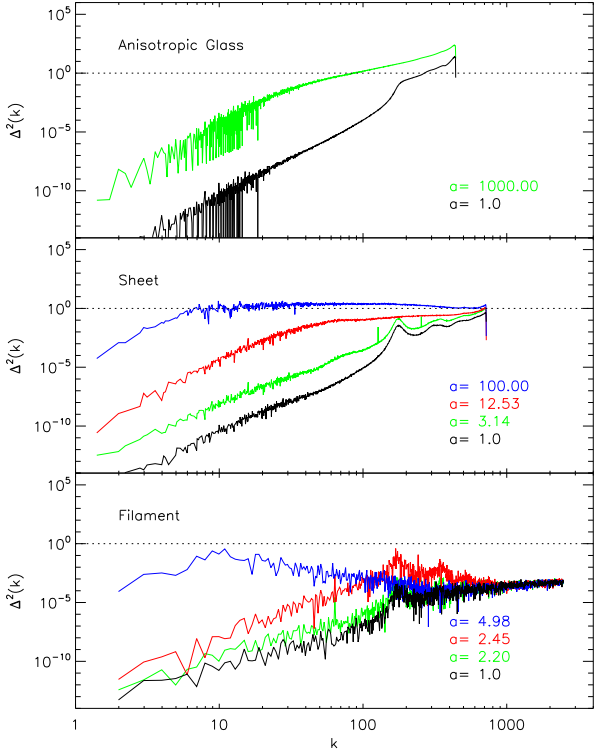}}\\%
\caption{ The evolution of the power per unit $\ln k$ in three idealised
simulations. Top: the 3-D power at the initial time and after expansion by a
factor of 1000 for an anisotropically compressed but space-filling
glass. Centre: the 2-D power at the initial time and after expansion by
factors of 3.1, 12.5 and 100 for the 1-D collapse of a compressed glass to a
sheet. First collapse occurs at $a=3.1$. Bottom: the 1-D power at the initial
time and after expansion by a factors of 2.2, 2.4 and 5 for the 2-D collapse
of a compressed glass to a filament. Here first collapse occurs at $a=2.2$. 
\label{fig5}}
\ec
\end{figure}

%%%%%%%%%%%%%%%%%%%%%%%%%%%%%%%%%%%%%%%%%%%%%%%%%%%%%%%%%%%%%%%%%%%%
%%%%%%%%%%%%%%%%%%%%%%%%%%%%%%%%%%%%%%%%%%%%%%%%%%%%%%%%%%%%%%%%%%%%

Whether these features can indeed explain the unexpected fragmentation
of filaments in our HDM simulations depends, of course, on how they
are amplified as the particle distribution evolves. For the first
structures to collapse in an HDM universe, this evolution can be
idealised as a succession of three phases \citep{zeldovich70}. During
early nonlinear growth, the tidal field causes a locally anisotropic
flow which first reverses along a single preferred direction while
continuing to expand (although at different rates) along the two
orthogonal directions. In the second phase, collapse along the
preferred axis gives rise to a quasi-two-dimensional sheet-like
structure, a ``pancake''. Collapse along one of the other two axes
then produces a filament. Finally, material flows along filaments to
produce dark matter haloes at their intersections. These features are
all clearly visible in Fig.~1, although in practice the different
phases overlap and interact significantly. In Fig.~5 we show the
results from a set of idealised simulations of anisotropic collapse
designed to explore how discreteness noise grows for a glass initial
load during these various phases.

To illustrate structure growth in the first of the above phases, the top panel
of Fig.~5 shows the evolution of the total 3-D power per $\ln k$ for evolution
from an anisotropically compressed, but space-filling and otherwise
unperturbed glass.  (It is easier for us to simulate the isotropic expansion
of an anisotropically distorted glass than the anisotropic expansion of an
initially isotropic glass.) The $160^3$ ``good'' glass of Fig.~4 was here
compressed by a factor of 2 along the $x$-axis and a factor of 3 along the
$y$-axis, then replicated 6 times in order to tile the full simulation cube.
The exact periodicities introduced by this procedure are responsible for the
regular gaps in power visible at low $k$ in the initial power histogram
(the black curve in Fig.~5). It is interesting that the ``bump'' in power at
the discreteness scale is broader than in Fig.~4 and now stretchess from near
$k=160$, the interparticle separation of the {\it original} glass past $k\sim
300$, the interparticle separation of the compressed glass. Clumping during
expansion from this initial condition is extremely slow. The green curve shows
the power distribution after expansion by a factor of 1000. By this time the
matter has aggregated into small dense knots which typically contain 50
particles, but on larger scales the distribution remains almost uniform. At
long wavelengths the power has grown by about six orders of magnitude, just as
predicted by linear theory.  Thus the characteristic mass of the clumps grows
as $M_*\propto a^{6/7}$.  The amplification of discreteness noise is 
very weak during the first phase of anisotropic evolution from glass initial
conditions.

The second panel of Fig.~5 studies the growth of discrete noise during and
after collapse to a sheet. We compress our $160^3$ glass by a factor of 2
along one axis, leaving the other half of the simulation cube empty. This
initial condition collapses to a thin uniform sheet and thereafter remains
thin with the particles oscillating about the symmetry plane. The figure shows
the total 2-D power per $\ln k$ in the projection of particle distribution
onto this plane at four different times: the initial time, the moment when the
mass first collapses to a thin configuration ($a=3.1$), the moment the sheet
reaches minimum thickness for the second time ($a=12.5$) and a substantially
later time ($a=100$). In this case the power on large scales grows much faster
than in the previous case with $\Delta^2$ increasing approximately as $a^6$
rather than as $a^2$. At the time of first collapse, the power in the
discreteness peak has grown rather little, even though the power
on larger scales has already amplified substantially. By the time of second
collapse many nonlinear clumps are already evident in the projected mass
distribution and the feature at the scale of the initial interparticle
separation is no longer visible in the power spectrum. The characteristic
nonlinear scale is determined by the point where the amplified long wavelength
$\Delta^2\propto k^6$ tail crosses $\Delta^2\sim 1$. This scale increases
rapidly with time, approximately as $M_*\propto a^3$. Structure in a pancake
thus grows by an accelerated version of the standard hierarchical aggregation
mechanism illustrated in a more familiar context in the top panel of
Fig.~5. Discreteness effects do not appear to play a role other than by
setting the initial amplitude of the long wavelength tail of $\Delta^2(k)$.

The lowest panel of Fig.~5 shows similar data for an idealised simulation of
collapse to a filament. We compress our $160^3$ glass by a factor of 2 along
two orthogonal axes, leaving the remaining three quarters of the simulation
cube empty. This bar-like initial condition collapses to a thin, straight
filament. The figure shows the 1-D power per $\ln k$ for the projection of the
particle distribution onto the axis of the filament at four different times:
the initial time, the time of first collapse ($a=2.2$), a time shortly
thereafter ($a=2.4$) and a significantly later time ($a=5$). The power spectra
here are considerably noisier here than in the top two panels because there
are far fewer modes per bin in $\ln k$. By first collapse the large-scale
power has grown substantially but there is rather little amplification near
the discreteness peak at $k\sim 160$. This is similar to the sheet
case. Shortly after first collapse, however, the power in the discreteness
peak has grown by a large factor, reaching nonlinear levels.  This shows up as
regular clumping along the filament with a periodicity close to $k=160$.  It
differs from the behaviour in the sheet case and is apparently analogous to
the filament fragmentation we saw in our HDM simulations. We investigate it
further in the next section. After the filament collapses the large-scale tail
of the power distribution amplifies extremely rapidly, roughly as
$\Delta^2\propto a^{20}$. At the last time plotted this growth is again
setting the nonlinear scale, as in the upper two panels, and there is no
obvious feature near $k=160$.

In all three of these tests the scale of nonlinearity at late times reflects
the amplified small-$k$ tail of the initial power spectrum of the glass.  This
tail grows much more rapidly in a sheet than in a uniform 3-D distribution and
much more rapidly in a filament than in a sheet. The initial glass, if well
made, does exhibit the theoretical minimum power on large scales
$P(k)\propto k^4$ (Zel'dovich 1965; Peebles 1980), so that no better
suppression of discreteness effects can be hoped for. In the filament case,
however, the {\it first} nonlinear structures are clearly different in nature
and are related to the interparticle separation scale of the {\it
uncompressed} glass.  We now consider this instability more closely.

%%%%%%%%%figure 6%%%%%%%%%%%%%
\begin{figure*}
\bc
\hspace{-1.6cm}
\resizebox{17cm}{!}{\includegraphics{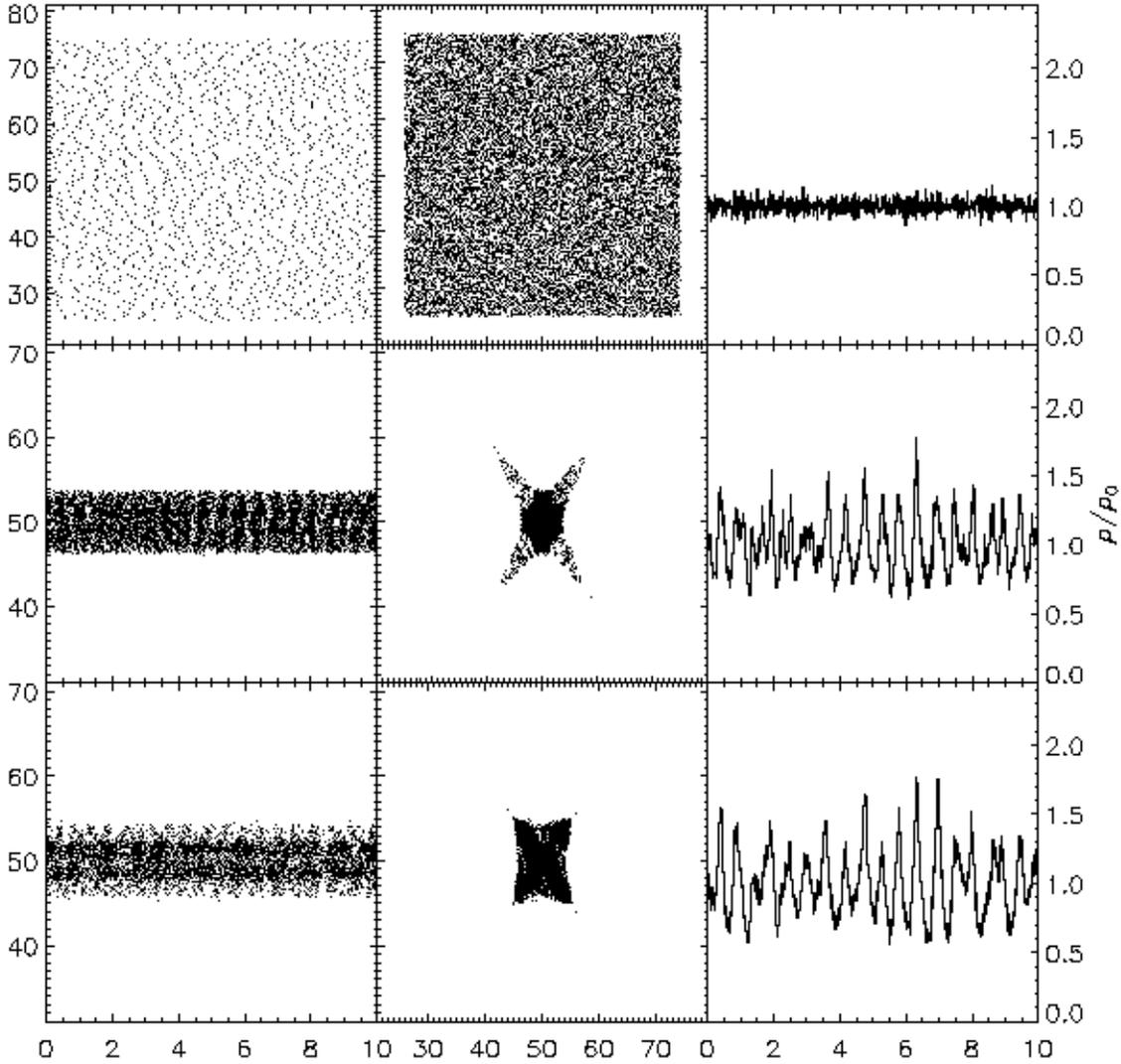}}\\%
\caption{The evolution of an idealised infinite straight filament. The first
and second columns are views perpendicular to and along the filament. Comoving
coordinates are here normalised so that the side of the computational cube is
100. The third column is the 1-D number density profile along the
filament. The density is normalised to unity and overdensities should be read
off against the scale on the right-hand-side of the plot. Results are shown
for expansion factors $a=1, 2.4$ and 2.6, from top to bottom. For clarity only
$1/10$ of the whole filament is shown here. In fact, the total number of
clumps lying along the filament is about 160.
\label{fig6}}
\ec
\end{figure*}

\section{Fragmentation of filaments}
\label{sec:filament}
In Fig. 6 we illustrate the evolution of the collapsing filament discussed in
the last section. The first and second columns show projections perpendicular
to and along the filament, while the third shows its 1-dimensional projected
density. Only a tenth of the full length of the filament is plotted in order
to make its structure more visible. Shortly after $a=2$ the filament collapses
to minimum thickness and at almost the same time it breaks up into regularly
spaced clumps. The clump spacing is very nearly equal to the mean
interparticle separation in the unperturbed glass; we find $\sim160$ clumps
along the full length of the filament.  The number 
of clumps is independent of the FFT grid used. For $128^3$, 
$243^3$, $400^3$ FFT grids, we find the total number of lumps to be
always around $\sim160$.  
Furthermore, we have repeated this fragmentation experiment with
different compression factors in the initial condition. This changes the time
of first collapse but it changes neither the fact that the filament breaks up
just after first collapse, nor the spacing of the clumps. The same is true
even if we adopt different compression factors along the two axes (provided
both are well above unity) or if we impose an initial perturbation which is
axially symmetric and has no sharp edges.  Using an initial glass with a
different number of particles produces a change in the interclump separation
which scales as the cube-root of $N$. Clearly then, the break up is associated
with a feature of the unperturbed glass.

The regular spacing of these artifacts indicates that modes with
$k\sim N^{1/3}$ dominate at least the early nonlinear evolution of
structure along the filament. This is visible in the left panels of
Fig. 7, which repeat the power spectra at the initial time and at
$a=2.4$ (just after collapse) from Fig.~5. At the initial time the
power around $k\sim 160$ is more than three orders of magnitude below
the threshold for nonlinearity, but by $a=2.4$ it is already approaching 
unity and is well above the power on all the other scales
plotted. This is the reflection in Fourier space of the remarkable
regularity seen in Fig.~6. The lower left panel of Fig.~7 plots growth
factors for individual modes between the two times.  The fastest
growing modes have $k$ somewhat smaller than 100, but their growth is
insufficient for them to overtake the initial power peak near $k\sim
160$. The power in all the modes in this peak grows by a similar
amount so the peak remains relatively narrow. This causes the
regular spacing of clumps along the filament.  Tests with a $270^3$
particle glass show identical behaviour but with the peak shifted to
$k\sim 270$. We conclude that the regularly spaced clumps which form
on the filaments of our HDM simulations are produced by a narrow peak
in power near the mean interparticle separation of our initial glass
load.  This peak is amplified to nonlinearity by the remarkably rapid
growth of structure which occurs once a filament has collapsed.
%%%%%%%%%%%%%%%%%%figure 7 %%%%%%%%%%%%%%%%%%%%
\begin{figure}
\bc
\hspace{-1.4cm}
\resizebox{8cm}{!}{\includegraphics{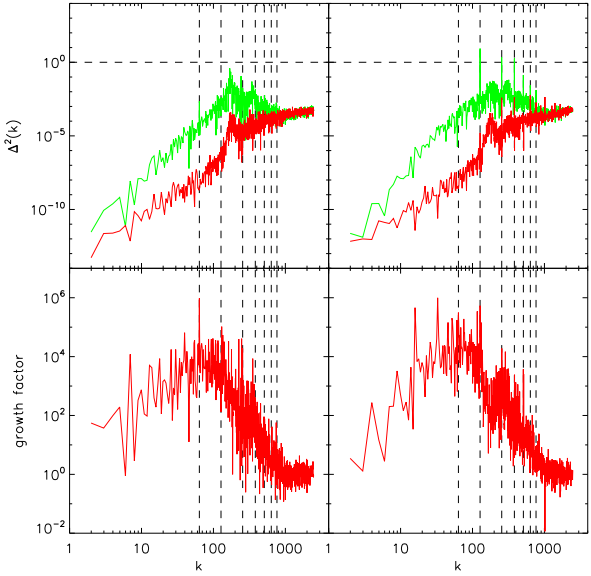}}\\%
\caption{ The top panels show the evolution of the 1-D power spectra
corresponding to the simulations in Fig.~6 (a ``good'' glass: left)
and Fig.~8 (a ``poor'' glass: right) which differ only in the initial
glass used.  The red curves represent the initial conditions ($a=1$)
while the green curves represent $a=2.4$.  The bottom panels show the
growth factors between the two times plotted mode by mode. Note that
these reach values in excess of $10^6$. The dashed black vertical
lines indicate modes $k=64, 128, 256, 384, 512, 640, 768$. It can be seen
that the growth factors are anomalously high for these modes in both
panels, and that in the case of the ``poor glass'' a number of them
are also anomalously high in the initial condition. As a result, at
the later time the structure along the filament is dominated in the
latter case by the modes with $k=128, 256$ and 384.
\label{fig7}}
\ec
\end{figure}

Careful examination of Fig.~7 shows that there are particular modes
for which the growth appears anomalously strong, notably those with
$k=64, 128, 256, 384...$. This is very likely a consequence of
anisotropies in Gadget's Poisson solver which is based on a binary
decomposition of the computational volume. For the ``good'' glass used
here these modes do not grow enough to overtake the power in the peak
associated with the interparticle separation, so it is the latter
which determines the initial fragmentation scale of the filament.
We now show that this is not always the case.

%%%%%%%%%figure 8%%%%%%%%%%%%%
\begin{figure*}
\bc
\hspace{-1.6cm}
\resizebox{17cm}{!}{\includegraphics{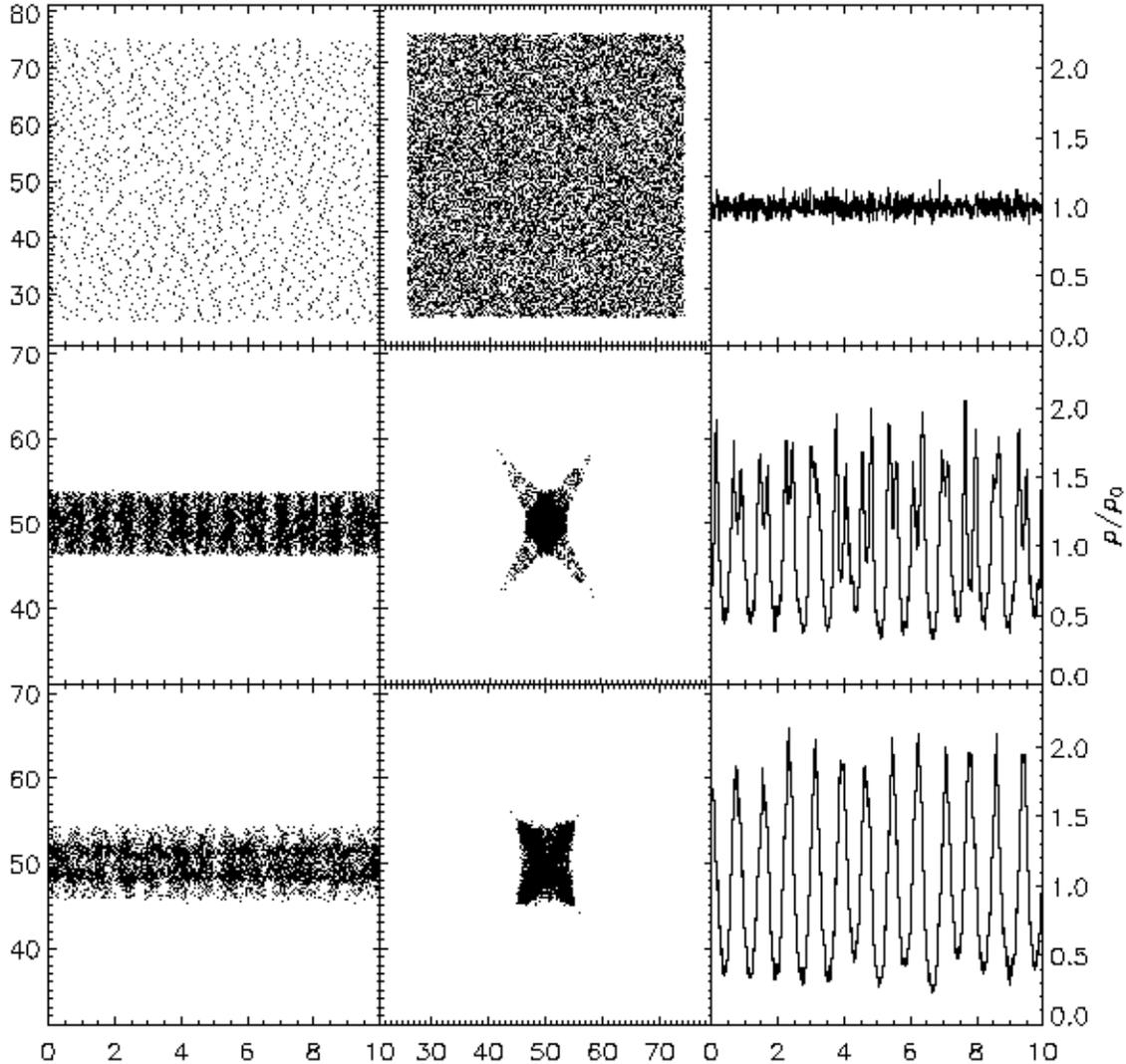}}\\%
\caption{Similar to Fig. 5. but starting from a ``poor'' glass where
the initial 1-D power spectrum shows spikes at $k=64, 128, 256..$.
There are about 128 clumps along the full length of the filament at the
last time shown.
\label{fig8}}
\ec
\end{figure*}

Up to this point all our results have been based on such a ``good''
initial glass for which the offset technique discussed in \S 2 was
used to minimise features due to anisotropies in the force
calculations.  Nevertheless, artifacts due to force anisotropies are
still visible in some of our plots. For example, spikes can be seen in
Fig.~5 at $k=128$ and 256 in the $a=3.1$ power spectrum of the sheet
(these spikes are solely due to modes with wave-vector parallel to the
fundamental axes of the computational cube) and at $k=64$ and 256 in
the $a=2.4$ power spectrum of the filament (see also the left panels
of Fig.~7). The growth factors in Fig.~7 show that these ``special''
modes grow substantially more rapidly than neighboring modes. If we do
not use random offsets to reduce the impact of algorithmic boundaries
in the force calculation, then features of this kind can be strong
enough in the initial glass to significantly affect later
evolution. An example is shown in the right panels of Fig.~7 and also
in Fig.~8. Here we have carried out exactly the same filament collapse
test as before, but using a ``poor'' initial glass constructed without
using the offset technique.  Spikes are now visible at $k=128, 256$
and $384$ in the {\it initial} 1-D power spectrum. These are amplified
by the evolution and at $a=2.4$ the power is dominated by the
amplified spike at $k=128$ rather than by modes in the neighborhood of
the discreteness peak at $k\sim 160$. Spikes at $k=256$ and 384 are
also very strong and several other spikes are clearly visible. As
Fig.~8 shows, these spikes cause the filament to break up initially
into $\sim 128$ rather than $\sim 160$ clumps.  Subsequent aggregation
into larger objects is similar in the two cases, however, with
large-scale effects overwhelming the initial differences.

Eliminating these troublesome power spikes from the initial conditions and
from subsequent evolution is not easy. Changing the nominal accuracy of the
force calculation affects the amplitude of the spikes but does not remove
them. We were surprised to find that similar spikes are present in the initial
glass used for the GIF simulations \citep{kauffmann99} even though this
was created using a different code based on a $\rm {P^3M}$ Poisson
solver. (For the GIF simulations the artifact was of no consequence, because
of the substantial small-scale power imposed in the CDM initial conditions.)
The large-scale PM force calculation in both codes imposes a regular ``power
of 2'' spatial structure, and for Gadget-2 this is reinforced by the static
Barnes-Hut oct-tree which underlies the calculation of the short range
forces. The unexpected spikes appear to reflect these structural properties of
the force construction algorithms. To test this, we projected the initial
glass onto periodic directions which are {\it not} aligned with the axes of
the computational box.  The corresponding 1-D power spectra do not show any
sharp spikes.

The tests in this section demonstrate that even with our random offset
procedure artifacts due to our Poisson solver are not entirely eliminated.  On
the other hand, these tests are extremely sensitive to such artifacts because
of the very high growth rates which occur in the idealised straight filaments
we have been studying. While it is clearly important to be aware of the
possibility of such numerical effects when simulating WDM or HDM universes,
our results here show that for a carefully constructed glass the effects due
to the Poisson solver remain subdominant with respect to effects caused by the
discreteness of the particle distribution. The latter cannot be eliminated for
any choice of initial particle load. They set the fundamental lower limit to
the effective resolution of such simulations
% %%%%%%%%%%%%%%%%%%%%%%%%%%%%%%%%%%%%%%%%%%%%%%%%%%%%%%%%%%%%%%%%%%
\section{Discussion and Conclusions}
\label{sec:conclusions}

%%%%%%%%%figure 9%%%%%%%%%%%%%
\begin{figure}
\bc
\hspace{-0.4cm}
\resizebox{9cm}{!}{\includegraphics{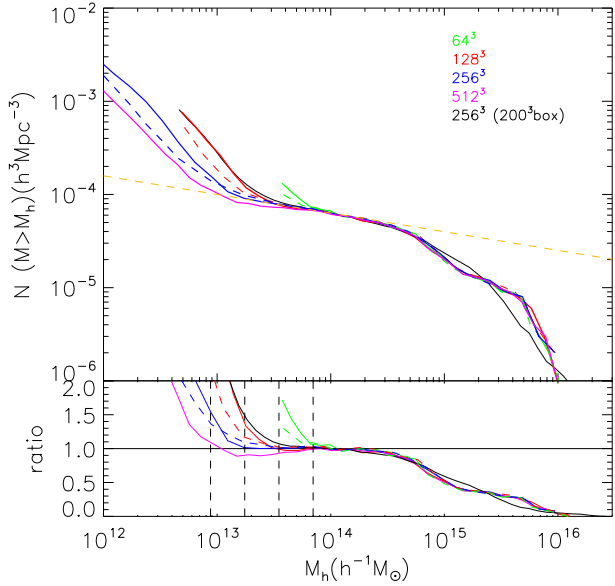}}\\%
\caption{ The cumulative halo mass function in eight simulations of an HDM
universe. Seven of these start from the same realisation of the HDM density
fluctuation field within a $100{\rm Mpc/h}$ box, but use different initial
particle loads. One follows evolution within a $200{\rm Mpc/h}$ box in
order to better constrain the high mass end of the mass function. Simulations
starting from a glass initial load are indicated by solid lines, while those
starting from a grid are indicated by dashed lines. The number of particles in
each simulation is indicated by labels in the upper panel. The dotted line in
this panel is an approximate power-law fit to the low-mass end of the mass
function,  $\rm N \sim M_{h}^{-0.2}$, and the lower panel
replots the mass functions relative to this
power law in order to emphasise the upturn due to discreteness effects.
Dashed vertical lines separated by factors of two provide a rough indication
of the scale where spurious halos start to dominate in the various cases.  The
haloes here were identified using an FOF algorithm with $b=0.2$ (Davis et al
1985).
\label{fig9}}
\ec
\end{figure}

In this paper we have studied how discreteness effects limit the effective
mass resolution of $N$-body simulations of cosmogonies like WDM or HDM where
structure on small scales is suppressed in the linear initial conditions.
Filaments occur in such models as part of the natural development of the
cosmic web, but in simulations they fragment into regularly spaced clumps with
a separation which reflects the mean interparticle distance in the initial
load. These spurious clumps are responsible for all the low-mass substructures
we have been able to identify at late times in collapsed halos.  Thus, it
appears that in an idealised WDM or HDM universe the first generations of dark
haloes are predicted to contain {\it no} self-bound substructures of
significantly smaller scale. Our tests on idealised systems show this 
fragmentation to occur because 1-D projections of a 3-D quasi-uniform 
particle distribution retain substantial power on the scale of the 
{\it 3-D} interparticle separation, and this power amplifies very
rapidly as the effectively 1-D system evolves.

We find (in disagreement with \citet{gotz02,gotz03}) that spurious
fragmentation of filaments occurs in almost identical fashion whether the
initial particle load is a glass or a grid.  Indeed, as we illustrate in
Fig.~9, the effect appears slightly worse for a glass than for a grid. This
plot gives mass functions for 9 simulations from HDM initial conditions which
use different particle numbers and different initial loads. In each case there
is a sharp upturn in abundance at small masses which reflects the clumps
visible within filaments in Figures 1 to 3. For initial loads of a given type
this upturn shifts to smaller masses by a factor of 2 for each factor of 8
increase in the number of particles. For a given number of particles the
upturn clearly occurs at somewhat larger masses in the glass case than in the
grid case. Notice also that the upturn for the $N=256^3$ glass simulation in a
$200h^{-1}{\rm Mpc}$ box agrees very well with that for the $N=128^3$ glass
simulation in a $100h^{-1}{\rm Mpc}$ box. This confirms that it is the mean
interparticle separation which sets the mass scale, rather than properties of
the simulation code or of the particular HDM realisation simulated.

If we take the effective lower resolution limit of our HDM simulations
to be given by the dashed vertical lines in the lower panel of Fig.~9,
we find that it can be expressed as $M_{lim} = 10.1 \times
\bar{\rho}~d~k_{peak}^{-2}$, where $\bar{\rho}$ is the mean density of
the universe, $k_{peak}$ is the wavenumber at the maximum of
$\Delta^2(k)$, the dimensionless power per $\ln k$ in the linear
initial conditions, $d= N^{-1/3}L$ is the mean interparticle
separation, $N$ is the number of simulation particles, and $L$ is the
side of the computational box. For our HDM initial conditions
$k_{peak} = 4.23\times\lambda_{fs}^{-1} = 0.1\times(m_\nu/30eV) 
{\rm Mpc}^{-1}$. The coefficient in our expression for $M_{\rm lim}$ 
is estimated directly from our HDM results. It may depend
significantly on the shape of the primordial power spectrum and so 
need modification for WDM initial conditions. The scaling 
$M_{\rm lim} \propto N^{-1/3}$ should still hold
in this case, however.  Comparing our formula without modification to
the numerical results of \citet{bode01} using $k_{peak}= 1.0$ and
$0.5 ~{\rm Mpc}^{-1}$, as appropriate for their two WDM models, gives
$M_{\rm lim} = 3\times 10^{10}$ and $1.2\times10^{11}h^{-1}{\rm
M_\odot}$. These values agree well with the upturns 
in the mass functions which they plot in their Fig.~9. 

We can also compare with the WDM simulations presented by
\citet{knebe02,knebe03}. These authors followed \citet{bode01} in
arguing that small mass haloes form along filaments by top-down
fragmentation.  However, if we compare the mass functions they present
for three different simulations in Fig. 4 of \citet{knebe02} and
Fig. 3 of \citet{knebe03}, we find the upturn at low mass for the two
simulations with the same numerical resolution but different WDM
particle mass to occur at masses which differ by about a factor of 4,
while the upturn for the two runs with different numerical resolution
but the same WDM particle mass also occur at masses differing by a
factor of about 4. This is in roughly agreement with the scaling we
predict for $M_{\rm lim}$ and is unexpected for a physical (rather
than numerical) feature. We conclude that these results, as well as
those of \citet{bode01} are consistent with a spurious numerical
origin for the low mass halos in filaments similar to that we find in
our HDM simulations.  Furthermore, with our parametrisation of the
characteristic scale based on the wavenumber at the peak of
$\Delta^2(k)$, the dependence of the characteristic mass of the effect
on the overall shape of the power spectrum appears to be weak.

This effective resolution limit is unfortunate news for simulations of HDM and
WDM universes. In our highest resolution HDM model, for example, the $N=512^3$
glass simulation of a $100{\rm Mpc/h}$ box, the resolution limit is $M_{lim} =
8.8\times 10^{12}h^{-1}{\rm M_\odot}$, which corresponds to a clump of $4300$
simulation particles. Thus only halos with $5000$ particles or more can be
considered reliable. This is two or three orders of magnitude below the
masses of typical big halos in the simulation. Contrast this with simulations
of CDM universes where the positions, velocities and masses of haloes are
reasonably well reproduced even for objects with about 100 simulation
particles, giving a logarithmic dynamic range which is about twice as
large. Furthermore the effective dynamic range in halo mass increases in
proportion to $N$ for CDM simulations, but only in proportion to $N^{1/3}$ in
HDM or WDM simulations.

These results are interesting for the question of whether WDM models can
reproduce the observed properties of dwarf satellite galaxies in the Milky
Way. Available kinematic data for dwarf spheroidals suggest that they are
sitting in dark matter halos with maximum circular velocities of order 30 km/s
\citep[e.g.][]{stoehr02,kazantzidis04} corresponding to masses (for
an {\it isolated} object) of about $10^{10}{\rm M_\odot}$.  After discounting
the spurious low-mass halos, the mass functions shown in Fig.~9 of \citet{bode01} 
demonstrate that halos of such small mass are not expected for a
WDM particle mass of 175 eV and are still strongly suppressed relative to
$\Lambda$CDM for a mass of 350 eV.  We infer that WDM particle masses well in
excess of 500 eV will be necessary to produce ``Milky Way'' halos with
sufficient substructure to host the observed satellites. This is, however,
less stringent by a factor of several than constraints based on structure in
the Lyman $\alpha$ forest \citep{viel06}. It will be interesting to
carry out simulations of sufficient resolution to test whether the internal
structure of subhalos in a WDM universe is consistent with that inferred for
the halos of Milky Way dwarfs. The resolution limitations we have explored in
this paper imply that, although possible, this will be a major computational
challenge.
%%%%%%%%%%%%%%%%%%%%%%%%%%%%%%%%%%%%%%%%%%%%%%%%%%%%%%%%%%%%%%%%%%%

\section*{Acknowledgements}
We thank Volker Springel for help in devising a glass-making scheme
which suppresses Poisson-solver-induced power spikes. We thank Adrian
Jenkins for the suggestion to consider idealised bar collapses. We
thank both of them and also Liang Gao for a number of very useful
discussions of the material presented in this paper.
\bibliographystyle{mn2e}
\bibliography{hdm1_v2}

\appendix
\section {The Quaquaversal distribution}
\label{qset}
\citet{hansen07} have suggested using an initial particle load constructed
through a simple ``Quaquaversal'' tiling of space (sometimes known as a
Q-set). In particular, they suggested using this initial load for WDM
simulations. In this appendix, we show what happens if this distribution is
used instead of a grid or glass initial load in a number of the tests we have
studied in our paper.

A Q-set ``lives'' in a rectangular box with side ratio $1:1:\sqrt{3}$, and
requires a total particle number of the form $N=2\times8^n$ with $n$ an
integer. Periodic boundary conditions can be assumed on opposite faces of the
box to represent an infinite cosmological distribution. Fig.~A1 compares the
3-D power spectrum of such a distribution with that of our ``good'' $160^3$
glass. This Q-set has $n=7$, resulting in a total of $N=4194304$ particles,
and was set up using the codes provided by \citet{hansen07}. In order to
facilitate the comparison we shift $P(k)$ for the Q-set so that its white
noise level is at $160^{-3}$ and the mode number $k=1$ corresponds to a
wavelength 160 times the mean interparticle spacing. At all scales
significantly larger than the mean interparticle spacing, the power in the
Q-set lies well above that in the glass. The mean power declines approximately
as $k^{3.4}$ for small $k$ rather than as $k^4$, and there is substantial power
in a series of additional narrow peaks separated by factors of 2 in $k$.
%%%%%%%%%figure A1%%%%%%%%
\begin{figure}
\bc
\hspace{-1.4cm}
\resizebox{8cm}{!}{\includegraphics{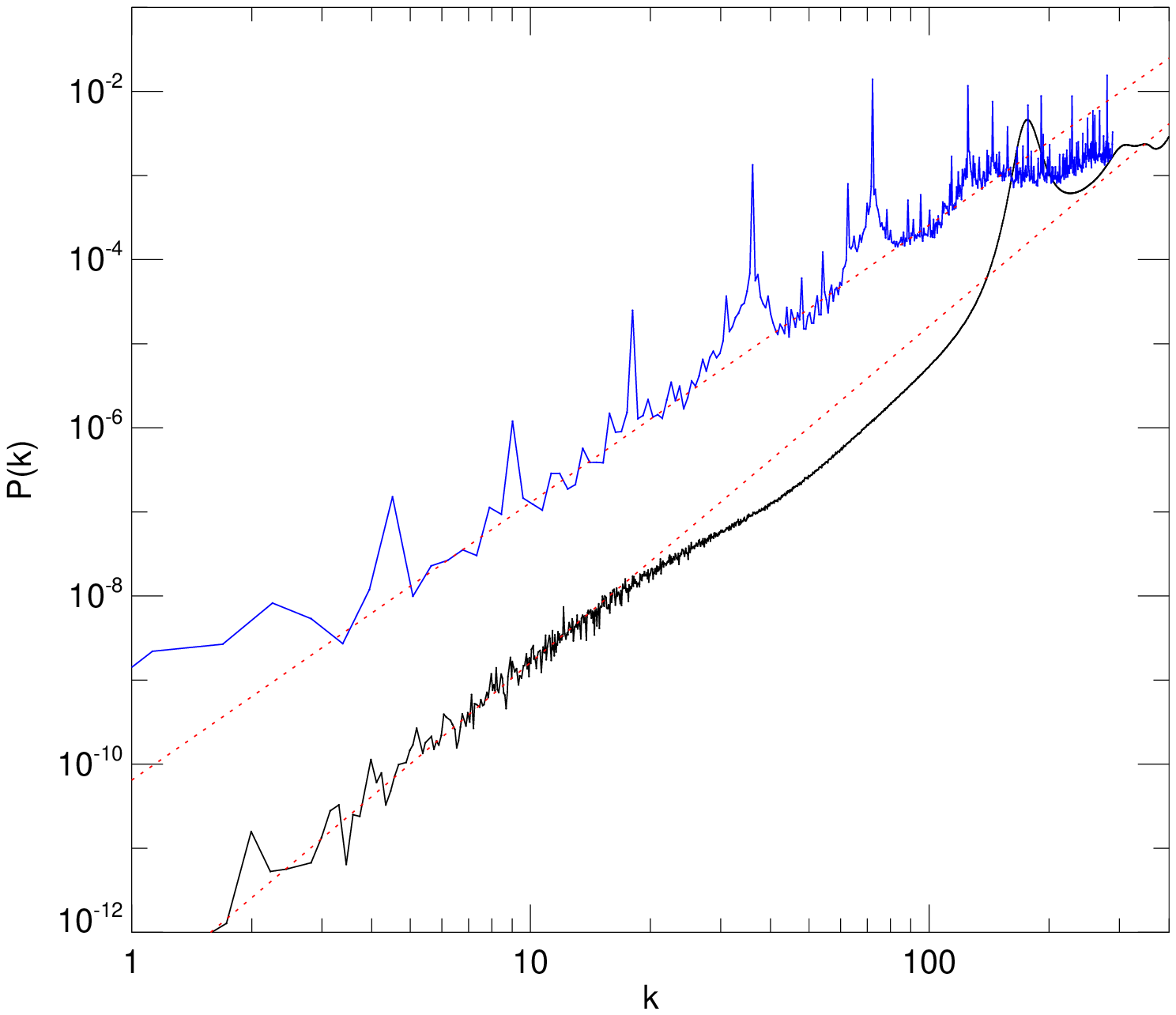}}\\%
\caption{3-D power spectra $P(k)$ for a Q-set (black) and for the ``good''
$N=160^3$ glass already used in Fig.~4 (blue).  The Q-set has 4194304
particles in a rectangular box. Its power spectrum is normalised so that its
white noise level is at $160^{-3}$, just as for the glass. In addition, the
units of wave number are chosen so that $k=1$ corresponds to a wavelength 160
times the mean interparticle separation for both distributions.  Red dotted
lines indicate power laws with slope 4 and 3.4 which are good fits to the
glass and Q-set power spectra respectively on large scales.
\label{figa1}}
\ec
\end{figure}

%%%%%%%%%figure A2%%%%%%%%
\begin{figure}
\bc
\hspace{-1.4cm}
\resizebox{8cm}{!}{\includegraphics{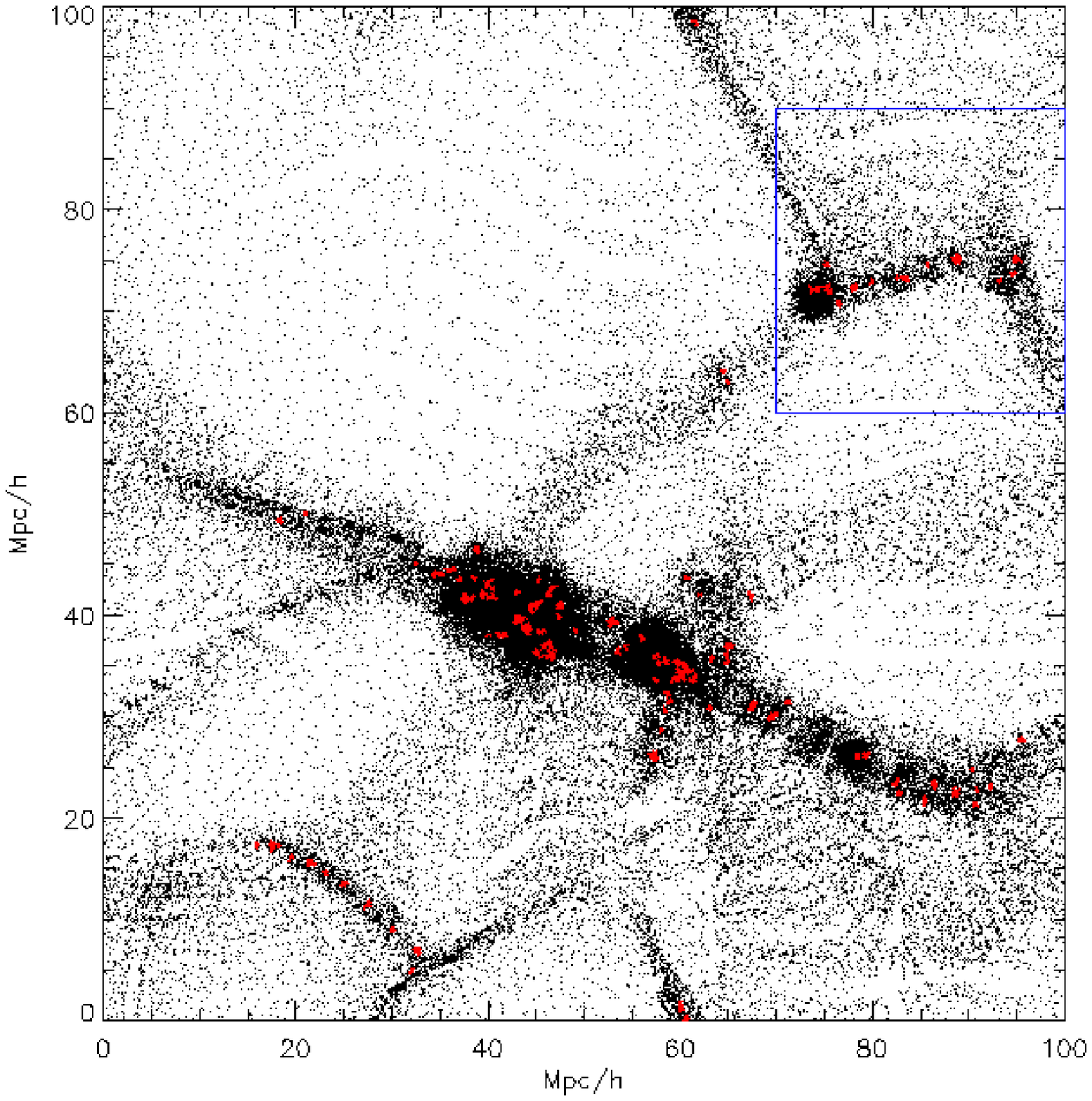}}\\%
\caption{A slice through an HDM simulation directly comparable to Fig. 1, 
but for a simulation with a Q-set initial load.
\label{figa2}}
\ec
\end{figure}

%%%%%%%%%figure A3%%%%%%%%
\begin{figure}
\bc
\hspace{-1.4cm}
\resizebox{8cm}{!}{\includegraphics{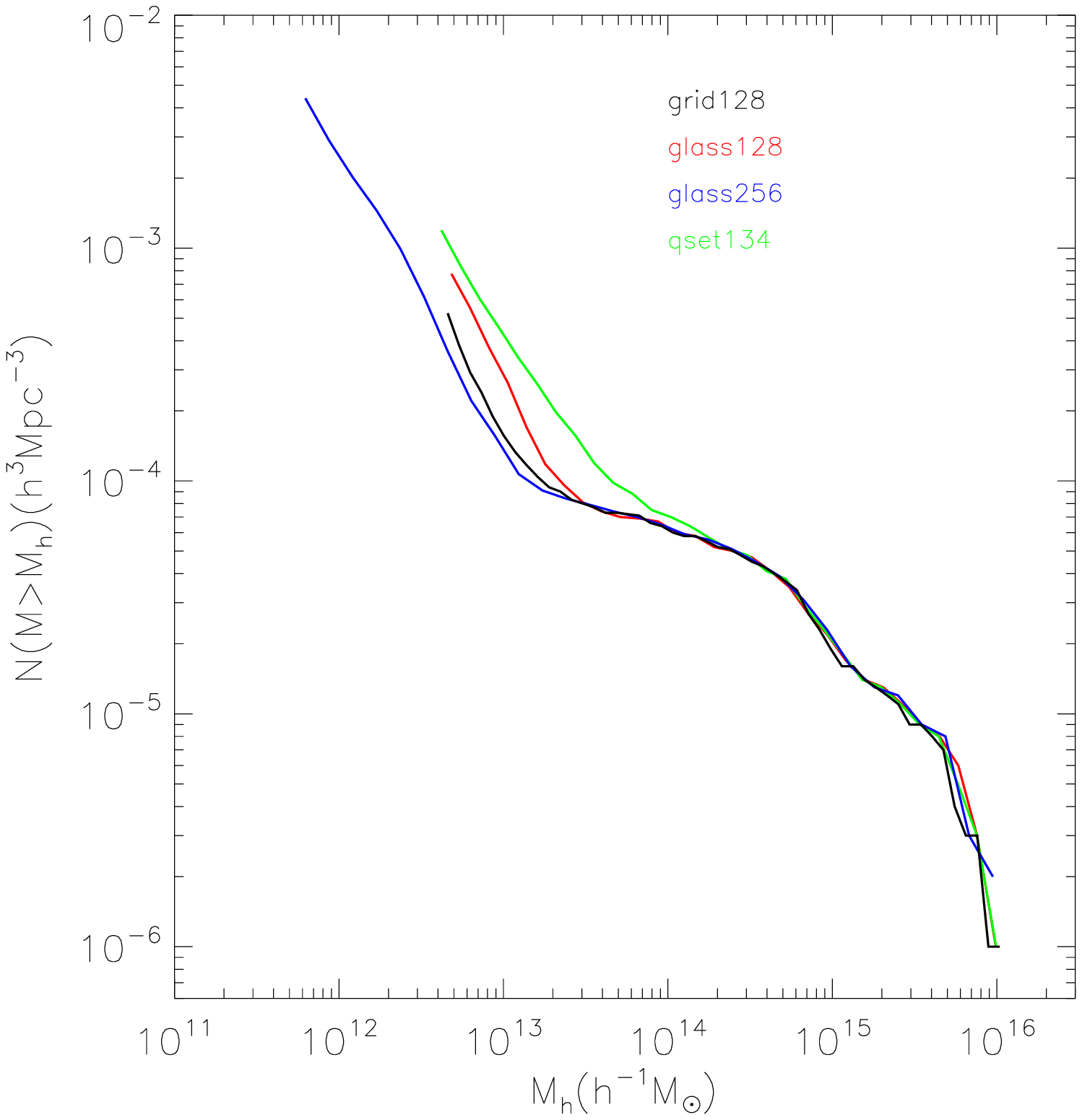}}\\%
\caption{Mass functions for FOF haloes in four HDM simulations started from
identical realisations of the HDM fluctuation field but with different initial
particle loads. Two start from glass loads with $N=128^3$ (red) and $N=256^3$
(blue) and one starts from a grid with $N=128^3$.  All three of these were
already shown in Fig.~9. The final simulation starts from a Q-set initial load
with $N\simeq134^3$.
\label{figa3}}
\ec
\end{figure}

We have used this Q-set as the initial particle load for our standard
$100h^{-1}{\rm Mpc}$ box HDM simulation. Since this simulation is carried out
in a cubic region we need to chop off the long end of the Q-set rectangular
box leaving a total of $2424140\sim 134^3$ particles in the simulation. This
truncation results in a violation of periodicity in the initial load for one
pair of faces of the cubic volume. As a result, spurious small clumps form
along this interface during later evolution, but these effects do not
propagate into the rest of the simulation. They are not visible in Fig.~A2,
which is a slice through the simulation to be compared directly with
Fig.~1.  We exclude a thin slice of the simulation which contains this
discontinuity when we calculate the halo mass function at $z=0$. This is
displayed in Fig.~A3 and compared to our earlier results.

%%%%%%%%%figure A4%%%%%%%%
\begin{figure}
\bc
\hspace{-1.4cm}
\resizebox{8cm}{!}{\includegraphics{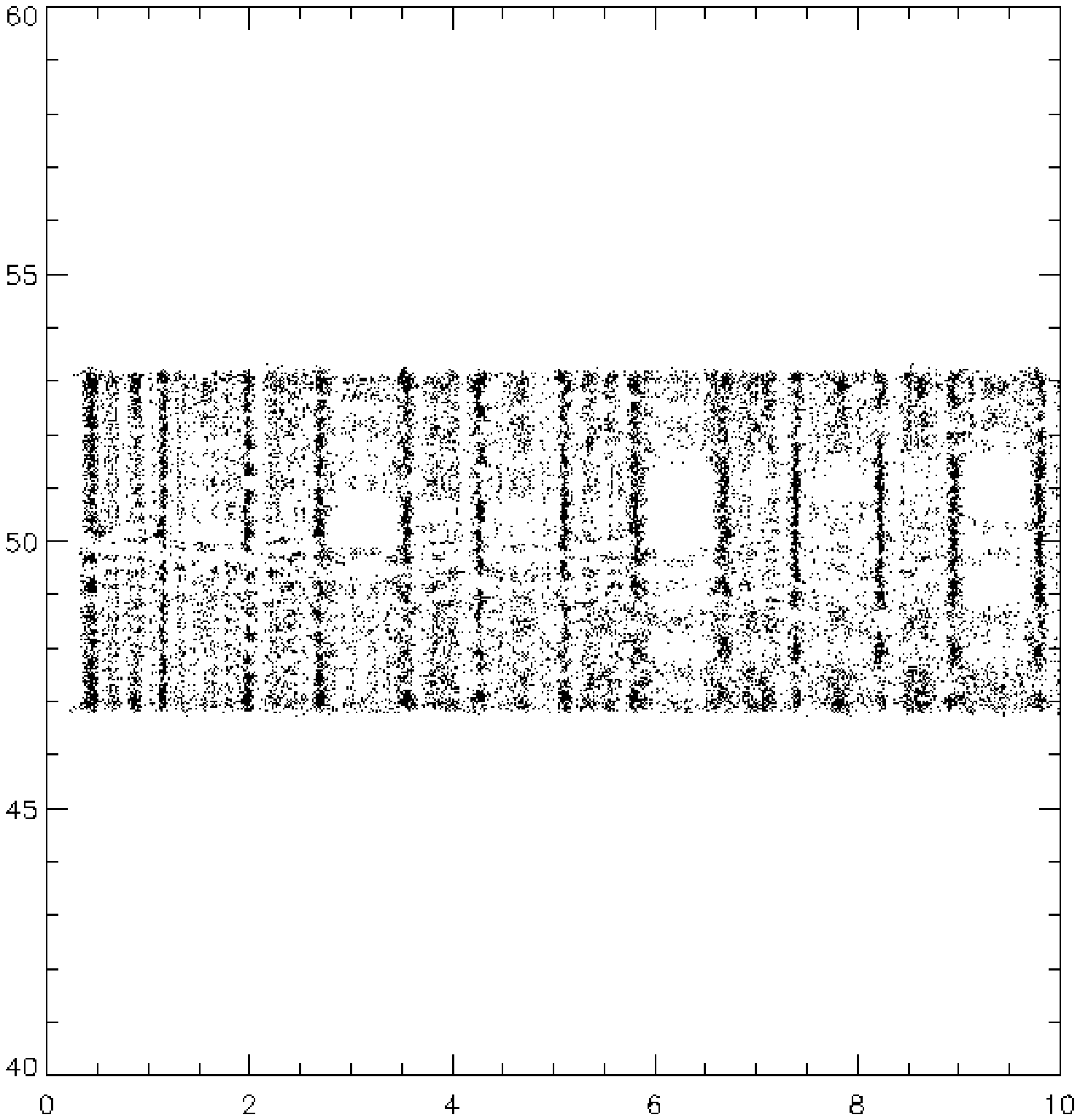}}\\%
\caption{Idealised filament collapse as in Fig.~6 but starting from a
compressed Q-set initial condition. The distribution of particles in a section
of the filament is shown at $a=2.2$ and should be compared directly with the
middle left panel of Fig.~6.
\label{figa4}}
\ec
\end{figure}

A comparison of Fig.~A2 with Fig.~1 shows that the regularities of the Q-set
are much more visually apparent than those of the glass, particularly in low
density regions. The small halos indicated by red points are again found along
filaments and on the outskirts of massive objects.  Figure A3 shows that the
Q-set reproduces the halo mass function of our other simulations at high
mass, but that the turn-up due to spurious low-mass halos occurs at
significantly higher mass than for glass or grid initial loads.  The effective
mass resolution of the Q-set simulation is about a factor of three worse than
in either of the other cases.
 
We have also carried out our standard idealised filament simulation
starting from a compressed Q-set rather than a compressed glass. The
particle distribution along a section of the filament is shown just
after collapse in Fig.~A4 for comparison with the central left-hand
panel of Fig.~6.  Substantial and regular clumping is seen, although
the regularity is considerably more complex (and also stronger) than
in the glass case.  As in the glass case, the clumps rapidly aggregate
into a small number of massive objects during later evolution.  The
different initial power spectra of the glass and Q-set cases
(Fig. A1), result in different rates of growth of structure along the
filament at later times. Structures are always more massive are
arranged in a more complex pattern along the filament for Q-set than
in the glass case.
 
Our conclusion from these tests is that for given particle number the Q-set
performs significantly worse as an initial load than either a grid
or a glass.  Since the visual regularities it induces are very similar to
those seen for a grid and are much stronger than those found with a glass,
there does not seem to be any obvious situation where a Q-set would be the
preferred choice for an initial quasi-uniform particle load.
 
\label{lastpage}
\end{document}